\documentclass[12pt,preprint]{elsarticle}
\usepackage[utf8]{inputenc}
\usepackage[boxed]{algorithm2e} 
\usepackage{amsmath}
\usepackage{amssymb}
\usepackage{enumitem} 
\usepackage{hyperref}
\usepackage{import} 
\usepackage[dvipsnames]{xcolor} 
\usepackage{booktabs}
\usepackage{siunitx}
\usepackage{mathtools}
\usepackage{scalerel}

\SetKwInput{KwInput}{Input}                                 
\SetKwInput{KwOutput}{Output}

\DeclareMathOperator*{\bigplus}{\scalerel*{+}{\sum}}

\newcommand{\BDY}{\phantom{ii}}
\newcommand{\whileC}{\mathop{\mathtt{while}}}
\newcommand{\doC}{\mathop{\mathtt{do}}}
\newcommand{\ifC}{\mathop{\mathtt{if}}}
\newcommand{\elseC}{\mathop{\mathtt{else}}}
\newcommand{\thenC}{\mathop{\mathtt{then}}}
\newcommand{\skipC}{\mathop{\mathtt{skip}}}
\usepackage{caption}
\usepackage{subcaption}

\begin{document}
\begin{frontmatter}

\title{Computable Model Discovery and High-Level-Programming Approximations to Algorithmic Complexity}

\author[posgrado]{Vladimir Lemus\fnref{funding1}}
\author[posgrado]{Eduardo Acuña\fnref{funding1}}
\author[posgrado]{Víctor Zamora\fnref{funding1}}
\author[ciencias]{Francisco Hernandez-Quiroz\fnref{funding2}}
\author[OIA,karolinska,turing,labores]{Hector Zenil\fnref{funding3}}

\address[posgrado]{
  Posgrado en Ciencia e Ingenieria de la Computacion, UNAM, C.U., 04510, CDMX, Mexico
}

\address[ciencias]{
Facultad de Ciencias, UNAM, C.U., 04510,
CDMX, Mexico
}

\address[OIA]{
Oxford Immune Algorithmics, RG1 3EU, Reading, UK
}

\address[karolinska]{
Algorithmic Dynamics Lab, Karolinska Institute, 171 77, Stockholm, Sweden
}

\address[turing]{Alan Turing Institute, British Library, NW1 2DB, London, UK
}

\address[labores]{Algorithmic Nature Group, LABORES, 76005, Paris, France
}

\fntext[funding1]{The authors were supported by {\sc conacyt} scholarships while working on this project. They also want to thank Luis Felipe Benítez Lluis for his help with the project.}

\fntext[funding2]{The author was supported by a sabbatical grant ({\sc dgapa-unam}) while working on this project.}

\fntext[funding3]{Corresponding author: hector.zenil@cs.ox.ac.uk}


\begin{abstract}
Motivated by algorithmic information theory, the problem of program discovery can help find candidates of underlying generative mechanisms of natural and artificial phenomena. The uncomputability of such inverse problem, however, significantly restricts a wider application of exhaustive methods. Here we present a proof of concept of an approach based on IMP, a high-level imperative programming language. Its main advantage is that conceptually complex computational routines are more succinctly expressed, unlike lower-level models such as Turing machines or cellular automata. We investigate if a more expressive higher-level programming language can be more efficient at generating approximations to algorithmic complexity of recursive functions, often of mathematical interest.
\end{abstract}

\begin{keyword}
Kolmogorov-Chaitin complexity, Algorithmic Information Theory, algorithmic randomness, generative model generation, IMP.
\end{keyword}

\end{frontmatter}

\section{Introduction}

Since first proposed as a measure of complexity \cite{kolmogorov}, approximations to, or measures inspired by, Kolmogorov-Chaitin or algorithmic complexity, have found many applications \cite{li:vitany}. But at its core, as an semi-computable measure, it can only be applied by indirect methods such as computable weaker versions based on statistical lossless compressibility \cite{li:vitany,cilibrasi:vitany} or computationally expensive approximations \cite{zenil:bdm, zenil:sub-turing}. Statistical compressibility is limited to what entropic measures can find not beyond trivial regularities and redundancies without resort to uncomputability \cite{review:complexity:zenil}. Alternative approximations by means of exhaustive searches are also very expensive \cite{Soler-Toscano:2014,zenil:nature} and are limited to small programs. For very small Turing machines this can be avoided through clever tricks that cannot be expanded to larger machines \cite{Soler-Toscano:2014} and often produce the kind of programs that humans would not find very interesting such as the ones generating highly recursive objects such as arithmetic or geometric progressions that mathematicians find of high interest~\cite{sloan}.  This kind of approach has inspired a whole line of research into computable models~\cite{algodyn} for causal and knowledge discovery~\cite{zenil:nature,zenil:iscience}.

In a seminal paper \cite{Calude-Stay:2008}, Calude and Stay explored non-terminating programs from a probabilistic viewpoint. In further papers \cite{Calude-Dumitrescu:2018,Calude-Dumitrescu:2020}, they also proposed techniques that can be applied to estimate the probability that a program will not halt after certain execution time.

This paper builds on Calude's and other people's work in order to combine exhaustive exploration of programs with an estimation of their non-halting probability to produce approximations of Kolmogorov-Chaitin complexity. We do this using a high-level programming language. IMP \cite{winskel} is used to investigate if a more expressive and efficient higher-level programming language can generate faster approximations to the algorithmic complexity of recursive functions, often of mathematical interest (e.g. short descriptions for arithmetic sequences).

Assuming optimality, we systematically generated all programs in the language IMP up to certain length and execute them in a controlled environment. We registered their output to find the smallest program to produce a given string, establishing its Kolmogorov-Chaitin complexity under this model. In order to deal with non-halting programs we estimated their halting probability and used it to execute those programs to ascertain whether they would stop within an acceptable threshold. 

\section{A Simple Imperative Programming Language}


For our programming language we decided on IMP, a simple textbook example of an imperative programming language \cite{winskel} that is Turing-complete. The syntax of IMP is given by the grammar in figure~\ref{fig:gram}.

\newcommand{\TT}[1]{\mbox{\tt #1}}

\begin{figure}[ht]
\begin{align*}
  P &\Coloneqq \skipC \mid X \mathop{\mbox{\tt :=}} A \mid \TT(P\mathop{\mbox{\tt ;}}P\TT) \mid \TT(\ifC B \thenC P \elseC P\TT) \mid
  \TT(\whileC B \doC P\TT) \\
  A &\Coloneqq N \mid X \mid \TT(A \mathrel{\mbox{\tt +}} A\TT) \mid \TT(A \mathrel{\mbox{\tt -}} A\TT) \mid \TT(A \mathrel{\mbox{\tt *}} A\TT) \\
  B &\Coloneqq \mathtt{true} \mid \mathtt{false} \mid \TT(A \mathrel{\mbox{\tt =}} A\TT) \mid \TT(A \mathrel{\mbox{\tt <}} A\TT) \mid \neg B \mid \TT(B\vee B\TT) \mid \TT(B \wedge B\TT) \\
  X &\Coloneqq \mathtt{x}\TT[N\TT] \\
  N &\Coloneqq \mathtt{0} \mid C \\
  C &\Coloneqq \mathtt{1}S \mid \mathtt{2}S \mid \dots \mid \mathtt{9}S \\
  S &\Coloneqq \mathtt{0}S \mid \mathtt{1}S \mid \mathtt{2}S \mid \dots \mid \mathtt{9}S \mid \epsilon
\end{align*}
\caption{IMP language grammar in Backus-Naur form}
\label{fig:gram}
\end{figure}


One can assume IMP has a straightforward structural operational semantics \cite{plotkin}. Registers of the type $x[N]$ hold natural numbers as possible values. But we are interested instead in single binary strings as the output of programs. We adopt the following convention to convert the values in the registers used by a program into binary strings by a two-step mapping:

\begin{enumerate}
\item \label{itm:mapping} For each register, we map its contents to a binary string according to the canonical ordering of strings (strings are first sorted by length and then ordered lexicographically).
\item \label{itm:concat} We take the concatenation of all the registers' strings to be the program's output. In other words, a program's output is computed by:
  \[\bigplus_{n=0}^{\infty}f(x[n])\]
  where $f(x[n])$ is the function that transforms $x[n]$ into a binary string.
\end{enumerate}


For step \ref{itm:mapping}, we can use the very efficient algorithm \ref{alg:stringmap}. This algorithm can be derived from the simple observation that strings of $0$s always correspond to numbers of the form $2^n-1$ and vice versa. This is because the first
string of $0$s corresponds to the number $1$ and there are always $2^n$ strings of size $n$. So, if the string corresponding to the number $2^n-1$ is the string of $n$ $0$s, then the string of $n+1$ $0$s must correspond to $2^n-1 + 2^n = 2^{n+1}-1$.

\begin{center}
\begin{algorithm}[H]
  \SetAlgoLined
  \KwInput{A natural number $n$}
  \KwOutput{The $n$th string in canonical order}
  $l \gets \lfloor\log_2 (n+1)\rfloor$\;
  $c \gets n - (2^l - 1)$\;
  \KwRet{\text{$c$ in $l$ bits}}
  \caption{Algorithm for obtaining the $n$-th string in canonical order.}
  \label{alg:stringmap}
\end{algorithm}
\end{center}

Step \ref{itm:concat} involves concatenating the strings in all registers. Since registers are initialised as $0$ and $0$ is mapped to the empty string, the concatenation will be finite, and will only involve those registers which were modified by a given program.

For a small example of how we compute a program's output, consider the following program:

\begin{center}
  \texttt{
  \begin{tabular}{rcl}
  (x[0] &:=& 2;\\
    (x[1] &:=& 1;\\
     x[2] &:=& 3))
  \end{tabular}}
\end{center}

At the end of the program's execution, register $0$ contains $2$, register $1$ contains $1$, register $2$ contains $3$, and the rest of the registers all contain $0$.

To compute the program's output, each register is converted to a string. So register $0$'s $2$ is converted to $1$, register $1$ is converted to $0$ and register $2$ is converted to $00$. The rest of the registers are converted to
$\epsilon$. Therefore the program's output is:

\[1+0+00+\bigplus_{i=3}^\infty\epsilon\ = 1000\]

As Kolmogorov-Chaitin complexity is concerned with the smallest program producing a certain output, we need a length metric. We will use the terminal symbols of the language such as {\tt skip}, {\tt while}, etc., the decimal digits, and so on to measure the length of a program.

A quick glance at our grammar shows that the insertion of parentheses after each construct guarantees the unicity of the syntax trees. It also yields prefix-free programs, which will be useful in a future extension of this project.

\section{Taming the Halting Problem}

\subsection{Workarounds for the Halting problem}

The Halting problem is the main reason why Kolmogorov-Chaitin's complexity is semi-computable. In order to know whether a certain program will output a given string at the end of its execution, we need to know whether the program will halt at all. But this is the paradoxical nature of an uncomputable problem \cite{turing}.

In~\cite{Soler-Toscano:2014}, the authors applied reduction techniques for Turing machines with 5 states and 2 symbols, including setting a time limit to run the machines, and were able to analyse the Kolmogorov-Chaitin complexity of a relatively large set of strings.

A small programming language, known as Brainfuck (BF), has also been used to show an empirical approximation~\cite{Cibej:2014} to Kolmogorov-Chaitin complexity, as well as in a first attempt to estimate a theoretical error measure for such approximations~\cite{Kohler:2005}.

The main idea behind the theoretical model of~\cite{Kohler:2005} is to trade strict algorithmic solutions for approximate solutions. In other words, a value is considered to yield a solution within a particular range of confidence. The central idea is to pass from the worst case scenario of non-halting to an average case. As a program either halts or loops forever, there is no such thing as average halting in the strict sense, but hard instances of the halting problem are rarely encountered in practice, according to empirical observations provided by Köhler, Schindelhauer and Ziegler (although this depends on the particular encoding of the problem).

Here we are using a completely different method based on the theoretical results of Calude and Stay~\cite{Calude-Stay:2008}. Their results hold for any universal Turing machine and posit an \emph{a priori} probability distribution over all running times. The probability space in this case is extended to both program lengths and runtimes, with time and programs uniformly distributed, in contrast to a previous approach by Chaitin based upon a probability space just over program length~\cite{Chaitin:1975}.

The central hypothesis is that long runtimes are in effect rare, and as such they tend to zero in all probability distributions. A method to decide when to stop a running program could be implemented by using an auxiliary Turing machine running a probabilistic \emph{anytime} algorithm~\cite{Grass:1996}. Runtimes for halting programs are algorithmically non-random, which means they can be generated by a machine with a small input, in this case the auxiliary Turing machine. 

Calude and Dumitrescu~\cite{Calude-Dumitrescu:2018} have further proved that it is possible to apply the above method to approximate a runtime limit for the overwhelming majority of halting programs. This limit depends solely on the machine codification and it is independent of size and number of programs analysed. Their method produces two intrinsically connected viewpoints, one probabilistic and the other statistical \cite{Calude-Dumitrescu:2020}.

Instead of proposing an \emph{a priori} probability distribution, given that (a)~long runtimes imply that the halting probability tends to zero and that (2)~halting time values are algorithmically non-random, we are able to build a model for the halting probability based on running times of a finite set of programs executed on a universal Turing machine. In its turn, an upper bound for halting times can be estimated.

To calculate a limit for running times, the cumulative distribution function (CDF) and a $(1-\epsilon)$-quantile is used. Beyond the $(1-\epsilon)$-quantile value there is the upper $\epsilon$-tail in which the halting probability is negligible, so these values can be used as a first qualitative measure for our approximation. For a program $x$ running on the machine $U$, if the execution has not halted after the above-mentioned quantile value, we declare that the program will not halt, even if there are a few programs that still halt beyond this limit.

\subsection{Decision error, precision and confidence parameters}

Thus far we have a probabilistic device proposing that some programs under a certain optimal machine $U$ may not halt according to a value $\epsilon\in(0,1)$, the decision error. Achieving this result requires a probability distribution, indirectly based on the CDF and the quantile, and a random variable over time~\cite{Calude-Dumitrescu:2020}

\begin{equation}
  RT=RT_U : dom(U)\rightarrow T_U,
\end{equation}

\noindent where $dom(U)$ is the set of all the halting programs under $U$.  

Now the idea is to approximate the distribution by a long sequence of times. For this reason an Empirical Cumulative Distribution Function is defined as:

\begin{equation}
  ECDF_{RT,N}((RT_1(\mathbf{x}),...,RT(\mathbf{x})); t) = \frac{\#\{ 1\leq i \leq N : RT_i(\mathbf{x})\leq t\}}{N}, 
\end{equation}

\noindent with an $N$-dimensional \emph{time sampling space}~\cite{Calude-Dumitrescu:2020}. Working with these values and using Hoeffding's inequality~\cite{hoeffding}, two parameters are given: (a)~the precision parameter $\lambda$, with $0<\lambda<\epsilon$, and (b) the confidence parameter $\delta$, with $1-\delta \in (0,1)$. The first one is a bound for the approximation to the CDF by the ECDF; the second one is a confidence level, an index of how good our sampling is.

Having in hand three rational values that can be fine-tuned according to whatever levels of error, precision and confidence are desired in an approximation, a sampling size $N$ can be calculated~\cite{Calude-Dumitrescu:2020}:

\begin{equation}
  N\geq N(\lambda,\delta) = \left\lceil \frac{1}{2\lambda^2}\cdot ln\frac{1}{\delta}  \right\rceil,
  \label{ec:nsize}
\end{equation}

Sampling $N$ independent halting programs in this manner and 
obtaining their running times are the first steps. Thereafter the times can be sorted in increasing order. The maximum time the sampled programs run until halting will be the threshold $\mathbf{T}$. Any program that has not stopped by time $t \geq T$ will be considered to never halt, with an error equal to $\epsilon$.

Two more approaches exist to provide a threshold time $\mathbf{T}$. When it is too hard to sample exactly $N(\lambda,\delta)$ halting programs, it is easier to find an affordable sample of size $\tilde{N}$. Two values can be given, $\epsilon,\lambda \in (0,1)$ with $\lambda < \epsilon$, and the third one can be obtained by:

\begin{equation}
  \delta(\tilde{N},\lambda) \geq exp(-2\tilde{N}\cdot \lambda^2).
\end{equation}
  
The second approach is as follows: if neither a dynamic nor an affordable sample size is possible, an upper bound for time $\mathbf{T}$ can be set as an initial parameter, adding $\epsilon$ and $\lambda$ with similar properties as in the last case. A bigger sample of programs is used to obtain $dom(\mathbf{U})$, that is, all the programs halting in a time $t < \mathbf{T}$. Afterwards the size of the set $N(T)$ is calculated using the values $\epsilon$, $\lambda$ and $\tilde{N}=N(T)$ as input for the second approach. Both ways may entail a decrease in the confidence level.

\begin{table}[ht]
  \centering
  \begin{tabular}{@{}cccc@{}}
    \toprule
    $\epsilon$  & $\lambda < \epsilon$  & $\delta$  & $N(\lambda,\delta)$  \\
    \midrule
    $0.01$  & $0.009$  & $0.01$ & $\num{28427}$ \\
    $0.01$  & $0.001$  & $0.01$ & $\num{2302586}$ \\
    $0.01$ & $0.001$  & $0.001$ & $\num{3453878}$ \\
    $0.001$ & $0.0005$  & $0.001$ & $\num{13815511}$ \\
    \bottomrule
  \end{tabular}
  \caption{Sampling set size $N(\lambda,\delta)$ calculated for different $\epsilon$, $\lambda$ and $\delta$ values.}
  \label{tab:val}
\end{table}

The sampling size may not be directly dependent on $\epsilon$, our main error value, but it is a bound for the precision parameter $\lambda$. Very optimistic values can be proposed for $\epsilon$, $\lambda$ and $\delta$ to obtain the size of the sampling set. The table \ref{tab:val} shows some values obtained with equation \ref{ec:nsize}.

\subsection{Generating programs sorted by length}
\label{sec:genproglengths}

Once we calculated a time limit for the execution of programs of length 9 or less, we executed \emph{all} of them and observed their output (or gave up on them if the execution took longer than the estimated threshold for halting). But the systematic \emph{generation} of all programs up to a certain length is not an easy task. In a future paper, we will cover all the technicalities involved, which have to do mainly with the combinatorial explosion of programs produced by the grammar. Here we simply present an overview.

The space of programs is modelled as an enumeration, mapping non-negative integers called positions to abstract syntax trees corresponding to valid programs, such that every position is mapped to only one program and every valid program is mapped to only one position.

The enumeration used for this paper consists of all IMP programs sorted first by their length and then lexicographically. This enumeration is built in stages.

Starting from the context-free grammar of the language, a base enumeration is constructed where positions are mapped to programs based solely on the production rules of the grammar. This is done by breaking apart a position $k$ as the syntax tree of a program is built from top to bottom. The position determines what kind of tree must be constructed and in turn the kind of tree determines how the position must be broken apart into the positions corresponding to each of its children. This enumeration satisfies the property that given programs $p$ and $q$ at positions $i$ and $j$ respectively, if $q$ is a subprogram of $p$, then $j < i$.

For any reasonable measure for the length of programs, the length of a program is strictly greater than the length of any of its subprograms, however the aforementioned property isn't strong enough to guarantee that if $q$ is smaller than $p$ and $q$ isn't a subprogram of $p$ then $j < i$, which is one of main goals in constructing an enumeration sorted by length.

A simple metric to measure the length of programs is defined. The chosen metric counts the number of nodes in the syntax tree, and considers numbers with $d$ digits as constructs of length $d$.

Using the base enumeration and the length metric, a counting function is defined which maps non-negative integers $\ell$ to the number of programs of length~$\ell$. The metric informs the way a particular length can be partitioned for different kinds of programs, while the base enumeration works as a blueprint for how to navigate the program space. Table~\ref{tab:imp-count} shows the relationship between program length and number of programs in the IMP language, along with the cumulative sum of these counts.

\begin{table}[ht]
    \centering
    \begin{tabular}{@{}crr@{}} \toprule
        & \multicolumn{2}{c}{Count} \\
        \cmidrule(l){2-3}
        Length & Individual & Accumulated \\
        \midrule
        $0$ & $\num{0}$ & $\num{0}$ \\
        $1$ & $\num{1}$ & $\num{1}$ \\
        $2$ & $\num{0}$ & $\num{1}$ \\
        $3$ & $\num{3}$ & $\num{4}$ \\
        $4$ & $\num{104}$ & $\num{108}$ \\
        $5$ & $\num{2124}$ & $\num{2232}$ \\
        $6$ & $\num{35770}$ & $\num{38002}$ \\
        $7$ & $\num{546611}$ & $\num{584613}$ \\
        $8$ & $\num{7991176}$ & $\num{8575789}$ \\
        $9$ & $\num{114513832}$ & $\num{123089621}$ \\
        $10$ & $\num{1631934090}$ & $\num{1755023711}$ \\
        $11$ & $\num{23318957744}$ & $\num{25073981455}$ \\
        $12$ & $\num{335696750370}$ & $\num{360770731825}$ \\
        \bottomrule
    \end{tabular}
    \caption{Number of IMP programs by length.}
    \label{tab:imp-count}
\end{table}

The base enumeration and the counting function are used to build enumerations of programs for a given length, called fixed-length enumerations. If $c$ is the number of programs of length $\ell$, the corresponding fixed-length enumeration, or $\ell$-length enumeration, maps positions smaller than $c$ to all programs of length $\ell$. Fixed-length enumerations are finite, given that the number of programs of a fixed length is finite.

Finally, the canonically sorted enumeration is built by considering the fixed-length enumeration from smallest to largest length. This is achieved by keeping track of the program counts. Let $k$ be any position, and let $\ell$ be the greatest length such that the cumulative sum of program counts $\overline{c}$ is less than $k$. Then the corresponding program is at position $k-\overline{c}$ of the $\ell$-length enumeration.

The canonically sorted enumeration has the advantage of sorting programs in a predictable order. However, in contexts where the order is not as relevant, the base enumeration has a couple of advantages:
\begin{itemize}
    \item The algorithm to build a program for a given position is simpler and more efficient both in time and space.
    \item The current implementation includes the inverse process of calculating the position for a given program efficiently.
    \item Enumerations with different orderings or for subsets of the language can be easily derived.
\end{itemize}

[By the way, the techniques used to build the sorted enumeration can be applied to any other programming languages and metrics. The base enumeration is constructed in a declarative way by means of a combinator language based on Euclidean division and an \(n\)-tuple pairing function akin to Cantor's pairing function for pairs of natural numbers \cite{cantor_beitrag_1877}. The rest of the stages follow a manual process once a metric has been defined, and as long as the metric computes its value from a program's constituent parts and the mechanism for combining them, then this manual process can be adapted in a straightforward way.]

\subsection{Sampling the program space}

The sorted enumeration is used to navigate the space of programs. Starting from position zero, programs are generated and executed successively, with the guarantee that any programs generated afterwards will be of the same length or longer, which is convenient if the goal is to approximate their Kolmogorov-Chaitin complexity.

The results obtained from this process are collected and analysed after exploring all programs of each length. The space of programs of length at most $\ell$ is denoted by $\mathcal{S}_\ell$. In this project we focus our attention on $\mathcal{S}_9$, which consists of $|\mathcal{S}_9| = \num{123089621}$ programs.

Following the statistical approach of Calude et al for estimating running time, we used a sampling size of $N=\num{13815511}$ corresponding to the last row of table~\ref{tab:val}.

The sample must consist of programs that halt. In order to achieve this, a random number generator is used to sample uniformly distributed integers in the interval $[0, |\mathcal{S}_9|)$. 
Interpreted as positions in the sorted enumeration, these integers are effectively encodings of IMP programs in $\mathcal{S}_9$. After generating a program, it is executed with a large running time threshold $\mathbf{T}$ in order to determine if it can be considered part of the sample.

When choosing a threshold we must be careful to not set it too low, which would potentially exclude many halting programs from the sample. On the other hand, setting the threshold too high will affect the performance of the sampling process when executing non-halting programs. We decided to accept the performance loss by fixing $T=10^4$, as this value is quite large in comparison to the lengths of programs being executed. We believe it is a reasonable threshold to use during the sampling process.

For all sampled programs we recorded their position in the sorted enumeration, their length, their output strings, and their running time. Table~\ref{tab:imp-count-sample} shows how the sample is distributed by program length and the percentage corresponding to the total number of programs.

\begin{table}[ht]
    \centering
    \begin{tabular}{@{}crr@{}} \toprule
        & \multicolumn{2}{c}{Count} \\
        \cmidrule(l){2-3}
        Length & Sampled & Percentage \\
        \midrule
        $1$ & $\num{0}$ & $\num{0}\%$ \\
        $3$ & $\num{0}$ & $\num{0}\%$ \\
        $4$ & $\num{12}$ & $\num{11.54}\%$ \\
        $5$ & $\num{245}$ & $\num{11.53}\%$ \\
        $6$ & $\num{4091}$ & $\num{11.44}\%$ \\
        $7$ & $\num{62258}$ & $\num{11.39}\%$ \\
        $8$ & $\num{902471}$ & $\num{11.29}\%$ \\
        $9$ & $\num{12846434}$ & $\num{11.22}\%$ \\
        \bottomrule
    \end{tabular}
    \caption{Number of sampled programs by length.}
    \label{tab:imp-count-sample}
\end{table}

\begin{figure}[ht]
    \centering
    \caption{Running time distribution across the sample.}
    \label{fig:rtime-sample-01}
    \footnotesize
    \includegraphics[width={220bp}]{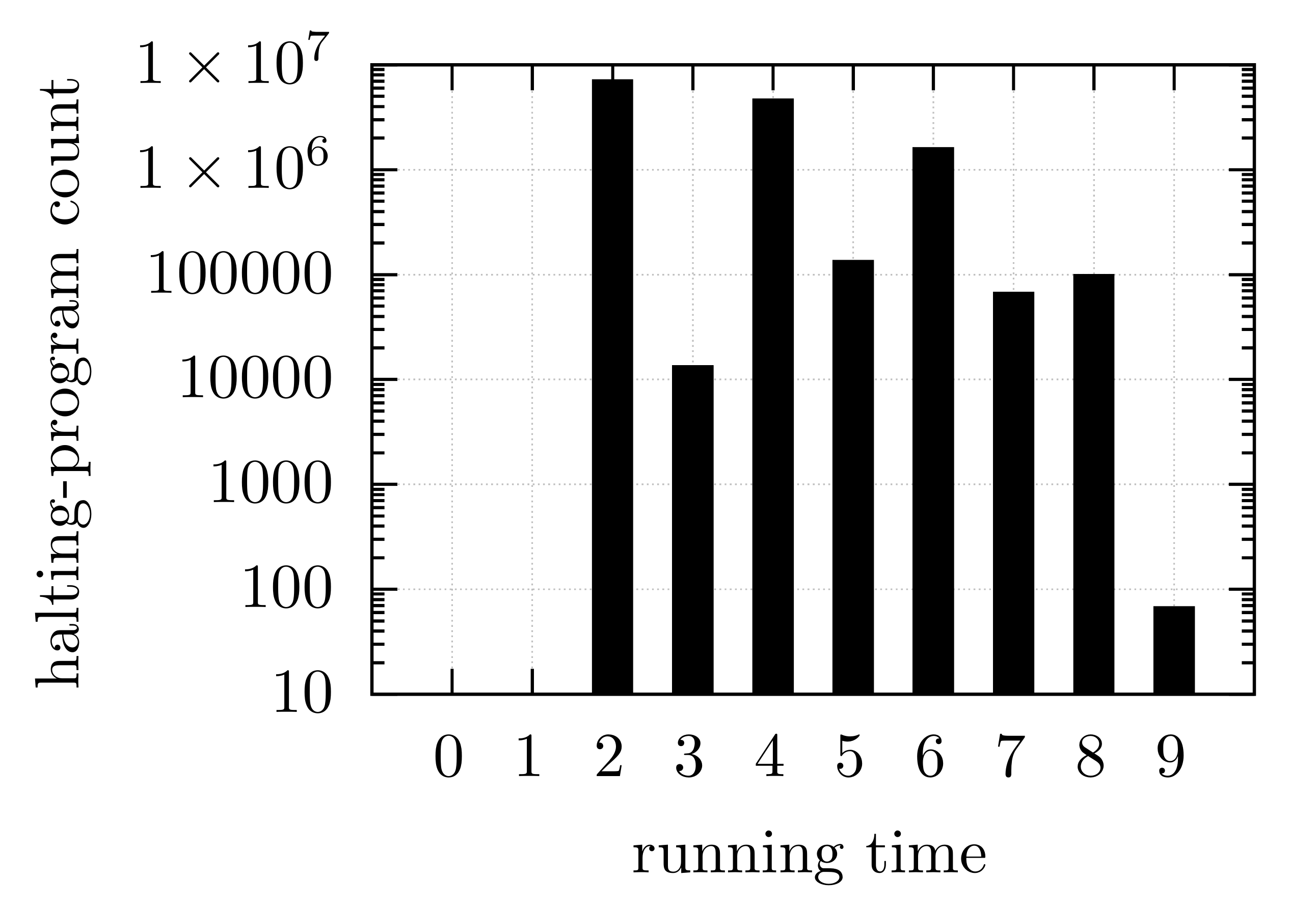}
\end{figure}

The largest running time obtained from the sampled programs was $9$ steps, which means that there were no sampled programs that halted between $10$ and $10^4$ steps, and indicates that the chosen running time threshold was high enough for our purposes. Figure~\ref{fig:rtime-sample-01} shows the distribution of the sample by the running time, where no programs ran in zero or one steps, and less than a hundred ran in $9$ steps.

\begin{figure}[ht]
    \centering
    \caption{Output length distribution across the sample.}
    \label{fig:output-sample-01}
    \footnotesize
    \includegraphics[width={288.00bp}]{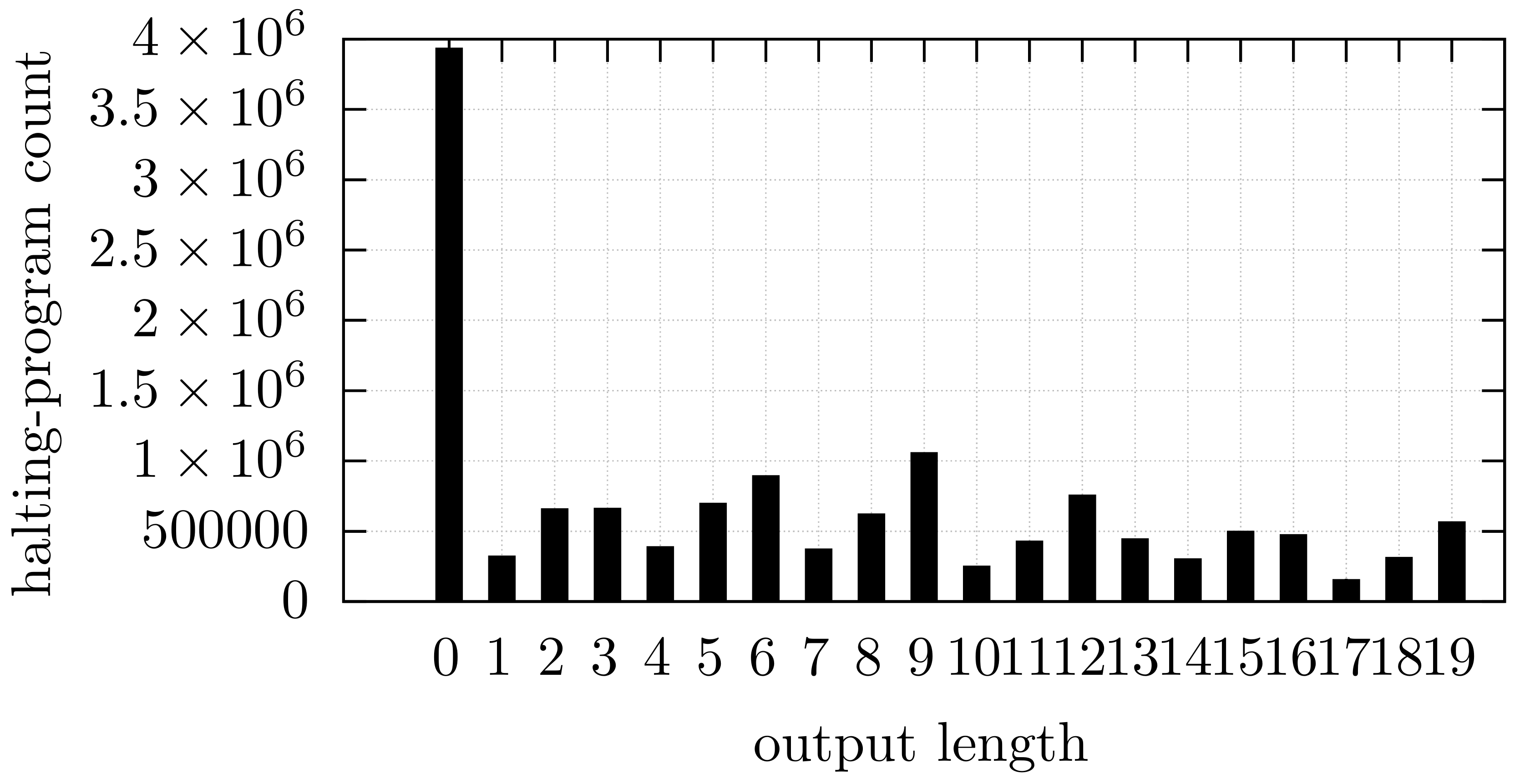}
\end{figure}

Figure~\ref{fig:output-sample-01} shows how the lengths of the outputs in the sample are distributed. Almost a third of the output strings obtained were the empty string, and the largest outputs consist of $19$ bits.

Given the results described above and the statistical estimation of Calude et al, we should run all programs in the sample space using a running time threshold of $\mathbf{T}=9$, which significantly reduces the evaluation time of non-halting programs, compared to the threshold $T=10^4$ used for the programs in the sample.

\section{Some results of the execution of programs}

\subsection{Systematic execution of programs in $\mathcal{S}_9$}

In order to empirically verify the quality of the chosen running time threshold estimate, we generated and executed all programs in $\mathcal{S}_9$ with threshold $\mathbf{T}$ to see if the statistically estimated $T$ is good enough.

Table~\ref{tab:length-halting} classifies the number of programs by length into halting and non-halting, along with the percentage of the total count. The results show a slight decreasing trend for the percentage of programs that halt as program length increases, and correspondingly the trend is inverted for non-halting programs.

\begin{table}[!ht]
    \centering
    \begin{tabular}{@{}crlrl@{}} \toprule
        Length & Halt & (\%) & Non-halt & (\%) \\
        \midrule
        $1$ & $\num{1}$ & $(\num{100.0}\%)$ & $\num{0}$ & $(\num{0.0}\%)$ \\
        $3$ & $\num{2}$ & $(\num{66.7}\%)$ & $\num{1}$ & $(\num{33.3}\%)$ \\
        $4$ & $\num{103}$ & $(\num{99.0}\%)$ & $\num{1}$ & $(\num{1.0}\%)$ \\
        $5$ & $\num{2059}$ & $(\num{96.9}\%)$ & $\num{65}$ & $(\num{3.1}\%)$ \\
        $6$ & $\num{34491}$ & $(\num{96.4}\%)$ & $\num{1279}$ & $(\num{3.6}\%)$ \\
        $7$ & $\num{522060}$ & $(\num{95.5}\%)$ & $\num{24551}$ & $(\num{4.5}\%)$ \\
        $8$ & $\num{7578748}$ & $(\num{94.8}\%)$ & $\num{412428}$ & $(\num{5.2}\%)$ \\
        $9$ & $\num{107800381}$ & $(\num{94.1}\%)$ & $\num{6713451}$ & $(\num{5.9}\%)$ \\
        \bottomrule
    \end{tabular}
    \caption{Halting and non-halting programs by length.}
    \label{tab:length-halting}
\end{table}

The table shows a sudden drop to $\num{66.7}\%$, which interrupts the trend mentioned above. This is because there are only three programs of length 3, and one of them is the first non-halting program: {\tt (skip; skip)}, {\tt(while false do skip)}, and {\tt(while true do skip)}.

The largest running time obtained from the sampling space was in fact $9$ steps, which confirms that the estimate $\mathbf{T}$ given by Calude's method would have given us the same results.

For the purpose of approximating the algorithmic complexity of a bit string, we are interested in finding non-trivial programs that produce an output longer that the programs' length.  However, for the purpose of finding generating computable models, any program of any length will provide, and the number of programs found for the same object, is an indication of the data's nature and its algorithmic probability determining a likelihood as a generating mechanism. 

\begin{figure}[ht!]
    \centering
    \caption{Distribution of $\mathcal{S}_9$ program length and running time.}
    \label{fig:length-rtime-log}
    \footnotesize
    \includegraphics[width={190.00bp}]{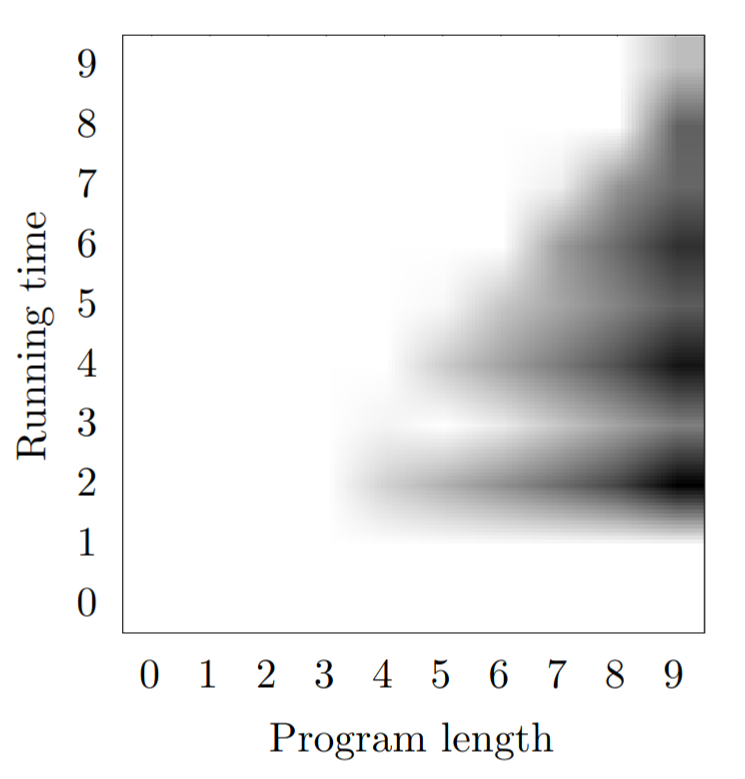}
\end{figure}

\begin{figure}[ht!]
    \centering
    \caption{Distribution of sample space by output length and program length.}
    \label{fig:output-length-log}
    \footnotesize
    \includegraphics[width={288.00bp}]{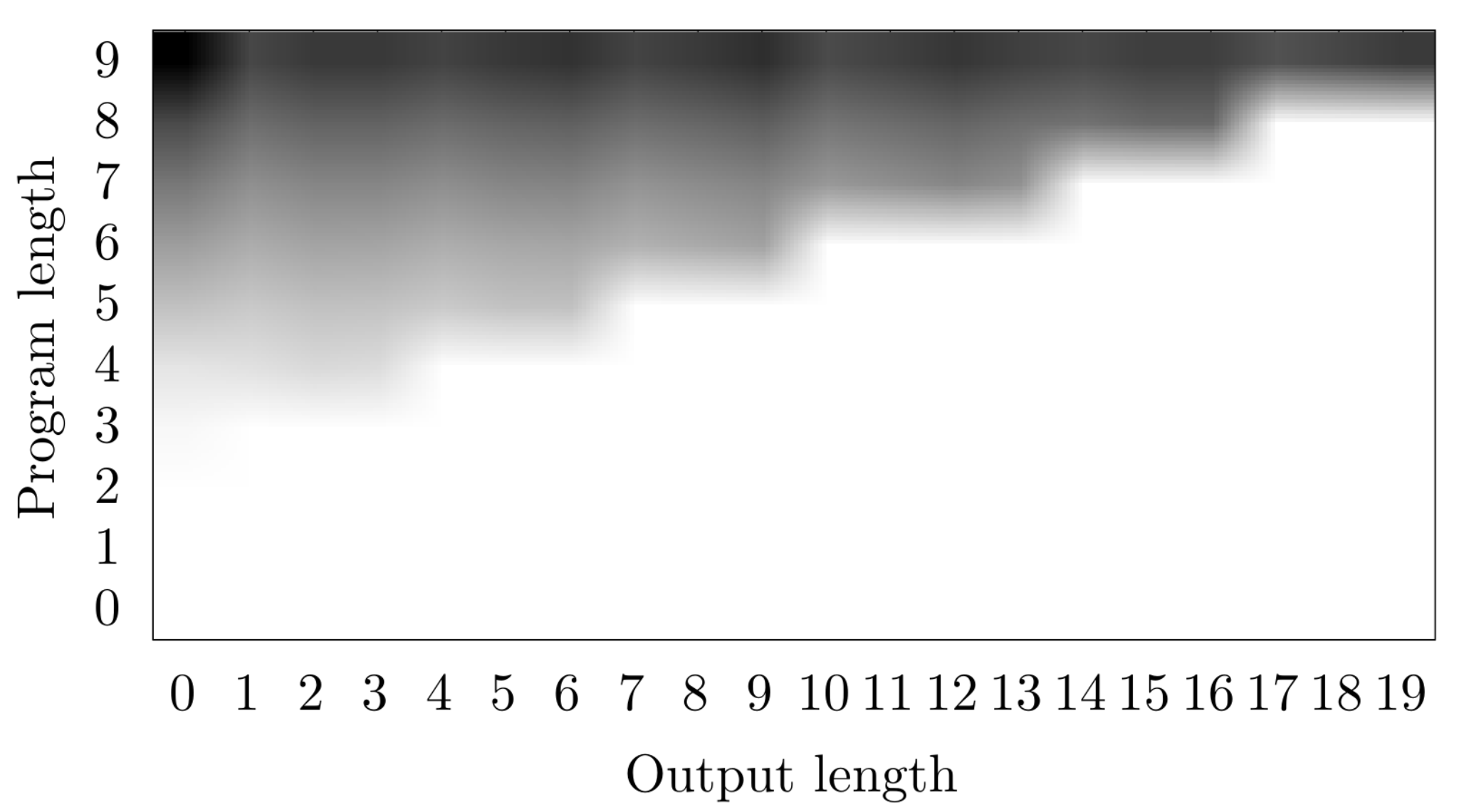}
\end{figure}

Figure~\ref{fig:length-rtime-log} shows the distribution of halting-programs by length and running time, and we expect non-trivial programs to appear in the upper-left triangle, where the running times are large relative to the length of the programs. Figure~\ref{fig:output-length-log} shows the distribution of halting-programs by output length and program length. We also expect non-trivial programs to appear in the lower-right triangle of figure~\ref{fig:output-length-log}, where the lengths of the output strings are large relative to the length of the programs.

In these figures, a darker shade means that there are more programs in the configuration--the intensity of the colouring is log-scaled in order to visualise clearly the distinction between zero programs and just a small number of programs.

In order to specify which programs are trivial in the context of Kolmogorov complexity, we use a standard process to calculate an upper bound for the complexity of a binary string $b$. First we compute the position $n$ of $b$ in the lexicographical enumeration of all binary strings. Then we build the program $p$ as $x[0] := n$. Finally the length $\ell$ of $p$ sets an upper bound on $K_{\mathrm{IMP}}(b)$. As a base case, we include the program $skip$ with length 1 as the upper bound for the empty binary string.

One future goal in exploring the program space is to improve upon this bound. Even though we explored around 123 million programs, and retrieved a million different output strings, the first smallest program found for every output string corresponded to the trivial program described above consistent with the theoretical expectation.

We found that the shortest programs produced a considerable number of strings. A table showing strings and a sample of the shortest programs producing them can be found at the the following URL:

{\small
\noindent\href{https://kolm-complexity.gitlab.io/approx-kolmogorov/approximating_kolmogorov.html}{https://kolm-complexity.gitlab.io/approx-kolmogorov/approximating\_kolmogorov.html}}

\subsection{Program space beyond $\mathcal{S}_9$}

While the systematic exploration of programs up to length 9 has allowed us to find the simplest programs producing some small strings, we are still at the level of trivial programs in the sense that their length is on the same order of magnitude than the strings they output. We need to go farther than $\mathcal{S}_9$ in order to find programs whose output is much bigger and therefore more interesting. In Figures~\ref{fig:progs-pows2}, \ref{fig:progs-fact}, \ref{fig:progs-expt}, \ref{fig:progs-expt-pows2} we present four families of programs producing an output clearly bigger than the length of the programs:

\begin{figure}[ht!]
    \centering
    \begin{minipage}{\textwidth}
        \ttfamily
        \begin{tabbing}
        (\=x[0] := 1;\\
        \>(while\ (x[1] < n) do\\
        \>\BDY(\=x[1] := (x[1] + 1);\\
        \>     \>x[0] := (x[0] * 2))))
        \end{tabbing}
    \end{minipage}
    \caption{Family of programs that calculate $2^n$.}
    \label{fig:progs-pows2}
\end{figure}

\begin{figure}[ht!]
    \centering
    \begin{minipage}{\textwidth}
        \ttfamily
        \begin{tabbing}
        (\=x[0] := 1;\\
        \>(while (x[1] < n) do\\
        \>\BDY(\=x[1] := (x[1] + 1);\\
        \>     \>x[0] := (x[0] * x[1]))))
        \end{tabbing}
    \end{minipage}
    \caption{Family of programs that calculate $n!$.}
    \label{fig:progs-fact}
\end{figure}

\begin{figure}[ht!]
    \centering
    \begin{minipage}{\textwidth}
        \ttfamily
        \begin{tabbing}
        (\=x[0] := 1;\\
         \>(while (x[1] < n) do\\
         \>\BDY(\=x[1] := (x[1] + 1);\\
         \>     \>x[0] := (x[0] * n))))
        \end{tabbing}
    \end{minipage}
    \caption{Family of programs that calculate $n^n$.}
    \label{fig:progs-expt}
\end{figure}

\begin{figure}[ht!]
    \centering
    \begin{minipage}{\textwidth}
        \ttfamily
        \begin{tabbing}
        (\=x[0] := n;\\
         \>(while (x[1] < n) do\\
         \>\BDY(\=x[1] := (x[1] + 1);\\
         \>     \>x[0] := (x[0] * x[0]))))
        \end{tabbing}
    \end{minipage}
    \caption{Family of programs that calculate $n^{2^n}$.}
    \label{fig:progs-expt-pows2}
\end{figure}

Examples of the output of these programs are shown in Tables~\ref{tab:progs-pows2}, \ref{tab:progs-fact}, \ref{tab:progs-expt} and \ref{tab:progs-expt-pows2}. The changing parameter is $n$. As it can see, for each family there is a lower bound on the value of $n$ when the output starts being clearly bigger than the length of the program itself.

\begin{table}[ht!]
    \centering
    \begin{tabular}{@{}rrrr@{}}
        \toprule
        & \multicolumn{2}{c}{Length} \\ \cmidrule(lr){2-3}
        $n$ & Program & Output & Output String \\
        \midrule
        $\num{0}$  & $\num{25}$ & $\num{1}$  & \texttt{0} \\
        $\num{1}$  & $\num{25}$ & $\num{2}$  & \texttt{10} \\
        $\num{2}$  & $\num{25}$ & $\num{3}$  & \texttt{011} \\
        $\num{3}$  & $\num{25}$ & $\num{5}$  & \texttt{00100} \\
        $\num{4}$  & $\num{25}$ & $\num{6}$  & \texttt{000101} \\
        $\num{5}$  & $\num{25}$ & $\num{7}$  & \texttt{0000110} \\
        $\num{6}$  & $\num{25}$ & $\num{8}$  & \texttt{00000111} \\
        $\num{7}$  & $\num{25}$ & $\num{10}$ & \texttt{0000001000} \\
        $\num{8}$  & $\num{25}$ & $\num{11}$ & \texttt{00000001001} \\
        $\num{9}$  & $\num{25}$ & $\num{12}$ & \texttt{000000001010} \\
        $\num{10}$ & $\num{26}$ & $\num{13}$ & \texttt{0000000001011} \\
        $\num{11}$ & $\num{26}$ & $\num{14}$ & \texttt{00000000001100} \\
        $\num{12}$ & $\num{26}$ & $\num{15}$ & \texttt{000000000001101} \\
        $\num{13}$ & $\num{26}$ & $\num{16}$ & \texttt{0000000000001110} \\
        $\num{14}$ & $\num{26}$ & $\num{17}$ & \texttt{00000000000001111} \\
        $\num{15}$ & $\num{26}$ & $\num{19}$ & \texttt{0000000000000010000} \\
        $\num{16}$ & $\num{26}$ & $\num{20}$ & \texttt{00000000000000010001} \\
        $\num{17}$ & $\num{26}$ & $\num{21}$ & \texttt{000000000000000010010} \\
        $\num{18}$ & $\num{26}$ & $\num{22}$ & \texttt{0000000000000000010011} \\
        $\num{19}$ & $\num{26}$ & $\num{23}$ & \texttt{00000000000000000010100} \\
        $\num{20}$ & $\num{26}$ & $\num{24}$ & \texttt{000000000000000000010101} \\
        $\num{21}$ & $\num{26}$ & $\num{25}$ & \texttt{0000000000000000000010110} \\
        $\num{22}$ & $\num{26}$ & $\num{26}$ & \texttt{00000000000000000000010111} \\
        $\num{23}$ & $\num{26}$ & $\num{27}$ & \texttt{000000000000000000000011000} \\
        \bottomrule
    \end{tabular}
    \caption{IMP programs computing $2^n$}
    \label{tab:progs-pows2}
\end{table}

\begin{table}[ht!]
    \centering
    \begin{tabular}{@{}rrrr@{}}
        \toprule
        & \multicolumn{2}{c}{Length} \\ \cmidrule(lr){2-3}
        $n$ & Program & Output & Output String \\
        \midrule
        $\num{0}$ & $\num{26}$ & $\num{1}$ & \texttt{0} \\
        $\num{1}$ & $\num{26}$ & $\num{2}$ & \texttt{00} \\
        $\num{2}$ & $\num{26}$ & $\num{2}$ & \texttt{11} \\
        $\num{3}$ & $\num{26}$ & $\num{4}$ & \texttt{1100} \\
        $\num{4}$ & $\num{26}$ & $\num{6}$ & \texttt{100101} \\
        $\num{5}$ & $\num{26}$ & $\num{8}$ & \texttt{11100110} \\
        $\num{6}$ & $\num{26}$ & $\num{11}$ & \texttt{01101000111} \\
        $\num{7}$ & $\num{26}$ & $\num{15}$ & \texttt{001110110001000} \\
        $\num{8}$ & $\num{26}$ & $\num{18}$ & \texttt{001110110000001001} \\
        $\num{9}$ & $\num{26}$ & $\num{21}$ & \texttt{011000100110000001010} \\
        $\num{10}$ & $\num{27}$ & $\num{24}$ & \texttt{101110101111100000001011} \\
        $\num{11}$ & $\num{27}$ & $\num{28}$ & \texttt{0011000010001010100000001100} \\
        \bottomrule
    \end{tabular}
    \caption{IMP programs computing $n!$}
    \label{tab:progs-fact}
\end{table}

\begin{table}[ht!]
    \centering
    \begin{tabular}{@{}rrrr@{}}
        \toprule
        & \multicolumn{2}{c}{Length} \\ \cmidrule(lr){2-3}
        $n$ & Program & Output & Output String \\
        \midrule
        $\num{0}$ & $\num{25}$ & $\num{1}$ & \texttt{0} \\
        $\num{1}$ & $\num{25}$ & $\num{2}$ & \texttt{00} \\
        $\num{2}$ & $\num{25}$ & $\num{3}$ & \texttt{011} \\
        $\num{3}$ & $\num{25}$ & $\num{6}$ & \texttt{110000} \\
        $\num{4}$ & $\num{25}$ & $\num{10}$ & \texttt{0000000101} \\
        $\num{5}$ & $\num{25}$ & $\num{13}$ & \texttt{1000011011010} \\
        $\num{6}$ & $\num{25}$ & $\num{17}$ & \texttt{01101100100000111} \\
        $\num{7}$ & $\num{25}$ & $\num{22}$ & \texttt{1001001000011111000000} \\
        $\num{8}$ & $\num{25}$ & $\num{27}$ & \texttt{000000000000000000000001001} \\
        \bottomrule
    \end{tabular}
    \caption{IMP programs computing $n^n$}
    \label{tab:progs-expt}
\end{table}

\begin{table}[ht!]
    \centering
    \begin{tabular}{@{}rrrr@{}}
        \toprule
        & \multicolumn{2}{c}{Length} \\ \cmidrule(lr){2-3}
        $n$ & Program & Output & Output String \\
        \midrule
        $\num{0}$ & $\num{26}$ & $\num{0}$ & $\varepsilon$ \\
        $\num{1}$ & $\num{26}$ & $\num{2}$ & \texttt{00} \\
        $\num{2}$ & $\num{26}$ & $\num{5}$ & \texttt{00011} \\
        $\num{3}$ & $\num{26}$ & $\num{14}$ & \texttt{10011010001000} \\
        $\num{4}$ & $\num{26}$ & $\num{34}$ & \texttt{0000000000000000000000000000000101} \\
        \bottomrule
    \end{tabular}
    \caption{IMP programs computing $n^{2^n}$}
    \label{tab:progs-expt-pows2}
\end{table}

Given that the shortest examples that show interesting results live in $\mathcal{S}_{25}$, it will take an enormous amount of computational work to generate and execute the programs in this class. In order to get the results in the tables we did not follow the canonical enumeration but the {\it base enumeration} mentioned in section \ref{sec:genproglengths}. Generating programs according to the base enumeration does not guarantee that we will find first the shortest program producing an output, but it can give us \emph{upper bounds} for the Kolmogorov complexity that are definitely smaller than the trivial programs. As this enumeration is also bijective and exhaustive it will eventually produce the shortest program too. In the meantime it has already given us useful information about the Kolmogorov complexity of strings representing the functions $2^n$, factorial, $n^n$ and $n^{2^n}$. The base enumeration can be regarded as a kind of \emph{look ahead} exploration.

\section{Conclusions and Future Work}

We started this project as a proof of concept for approximating Kolmogorov complexity using a very different computational model from those used in previous work. We managed to overcome some very serious technical impediments to efficiently generating vast amounts of programs, and we are confident the project can be scaled up for longer programs. But before discussing how this may be done, we want to draw some conclusions about the already explored space of programs.

First of all, we saw that Calude's method for estimating probable halting times can be applied to our chosen computational model without much trouble. Better still, estimating halting times for bigger program spaces is feasible (we have done this for programs of length 11 already, without need of considerably greater computational resources). 

\begin{figure}[ht]
    \centering
    \caption{Distribution of $\mathcal{S}_{11}$ by program length and running time.}
    \label{fig:length-rtime-log-11}
    \footnotesize
    \includegraphics[width={216.00bp}]{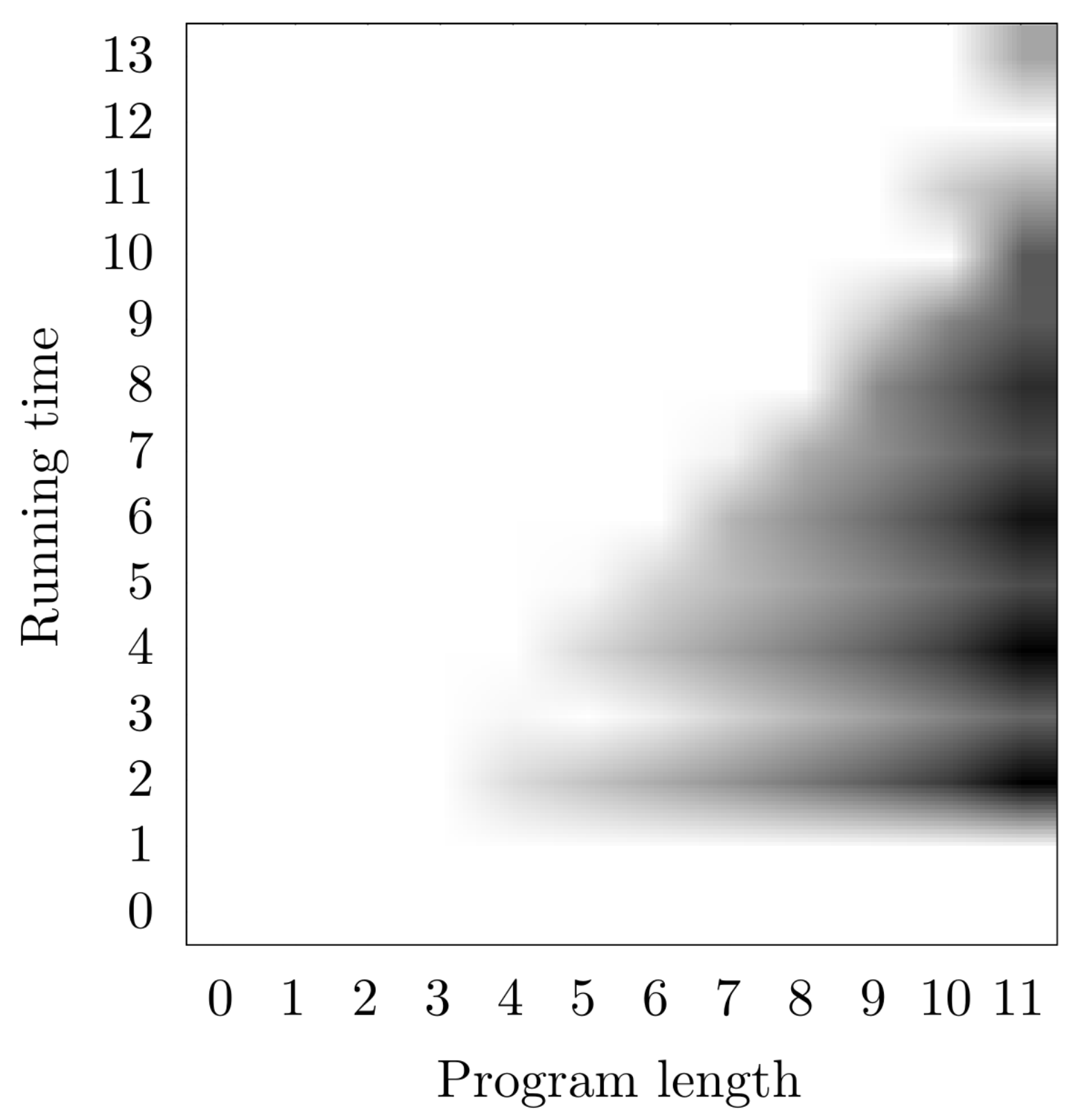}
\end{figure}

Secondly, all the programs found in $\mathcal{S}_9$ so far are still trivial, in the sense that their length is greater than that of the output strings. This was to be expected, as short strings can be described very succinctly by just presenting them. Yet this approach represents an opportunity to rank them even when short. Other approaches, including lossless statistical compression algorithms, present the same challenges and limitations, and are actually unable to rank short strings for more basic reasons. But there are indications that this could change for programs in $\mathcal{S}_{11}$. As this space is considerably larger than $\mathcal{S}_9$, we could not  collect the outputs produced by all programs in $\mathcal{S}_{11}$ that will be calculated in the future. Nevertheless, Figure~\ref{fig:length-rtime-log-11} shows the distribution of $\mathcal{S}_{11}$ by program length and running time. In contrast to figure~\ref{fig:length-rtime-log}, we are starting to get programs in the upper-left region above the identity line. Specifically, there are some programs of length 10 with a running time of 11 steps, and programs of length 11 with a running time of 13 steps, which means (quite significantly) that we are starting to see programs whose execution times exceed their length, and we think this region will grow faster in larger spaces. Longer execution times are needed to produce larger outputs.

Thirdly, using a different enumeration of programs we managed to explore much bigger spaces than $\mathcal{S}_9$ showing how the method can generate non-trivial programs and can find highly recursive ones quickly without many resources even at the beginning of the enumeration by choosing a high-level computer language.  This is key for the causal discovery program in~\cite{zenil:nature,algodyn} with applications to, for example, areas of Artificial Intelligence~\cite{frontiers}.

On the other hand, it is naive to think that such a vast and ever growing enterprise as calculating Kolmogorov complexity can be undertaken by a small team with limited resources. Currently, our workflow looks like this: we read a list of programs, run them, and save the outputs in a text file. After receiving the outputs, we check that the file is accurate through two mechanisms: an integrity check using hashes and a correctness check through the execution of a random sample of programs.

This process can easily be made to work within a collaborative framework. Among others, one option is to use the framework of the Automacoin\cite{noauthor_automacoin_nodate} cryptocurrency
tocken. A user of Automacoin  will get a reward for running millions of programs and submitting their output. This method of collaboration is sustainable and useful for both parties.

Collaborative frameworks need portable and efficient ways of setting and running partial tasks for collaborators. A portable, light-weight interpreter for IMP will be part of this effort. We are working on this and we already have a nice prototype implemented in Haskell \cite{noauthor_haskell_nodate}.

\end{document}
\endinput